\documentclass[12pt]{article}
\begin{document}
\title{Confined Maxwell Field and Temperature Inversion Symmetry}
\author{F. C. Santos$\thanks{%
filadelf@if.ufrj.br}$ and A. C. Tort$\thanks{%
tort@if.ufrj.br}$ \\
Instituto de F\'{i}sica \\
Universidade Federal do Rio de Janeiro\\
\\
Cidade Universit\'{a}ria, C.P. 68528 \\
\\
CEP 21945-970 Rio de Janeiro RJ, Brazil}
\date{\today }
\maketitle

\begin{abstract}
We evaluate the Casimir vacuum energy at finite temperature associated with
the Maxwell field confined by a perfectly conducting rectangular cavity and
show that an extended version of the temperature inversion
symmetry is present in this system.

\vfill

PACS: 11. 10. -z; 11. 10. Wx

\clearpage
\end{abstract}

\section{\protect\bigskip Introduction.}

Temperature inversion symmetry occurs in the free energy associated with the
Casimir effect \cite{Casimir1948} at finite temperature and it is linked to
the nature of the boundary conditions imposed on the quantum vacuum
oscillations. As shown by Ravndal and Tollefsen \cite{R&T89}, temperature
inversion symmetry holds for massless bosonic fields and symmetric boundary
conditions and also for massless fermionic fields and antisymmetric boundary
conditions. Temperature inversion symmetry was initially studied by Brown
and Maclay \cite{Brown&Maclay} who wrote the scaled free energy associated
with the Casimir effect at finite temperature as a sum of three
contributions, namely: a zero temperature contribution, that is, the Casimir
energy density at zero temperature, a Stefan-Boltzmann contribution
proportional the fourth power of the scaled temperature $\xi :=Td/\pi $, and
a non-trivial contribution. The three contributions to the scaled free
energy can be combined into a single double sum: 
\begin{equation}
f(\xi )=-\frac{1}{16\pi ^{2}}\sum_{n=-\infty }^{\infty }\,\sum_{m=-\infty
}^{\infty }\frac{(2\pi \xi )^{4}}{[m^{2}+(2\pi n\xi )^{2}]^{2}}\,,
\label{CONDPLATES}
\end{equation}
where the term corresponding to $n=m=0$ is excluded. Setting $n=0$ and
summing over $m$ with $m\neq 0$, we obtain the Stefan-Boltzmann limit $-\pi
^{6}\xi ^{4}/45$. On the other hand, setting $m=0$ and summing over $n$ with 
$n\neq 0$, we obtain the zero temperature Casimir term $-\pi ^{2}/720$. This
function has the following property: 
\begin{equation}
(2\pi \xi )^{4}f\left( 1/4\pi \xi \right) =f\left( \xi \right) ,  \label{TIS}
\end{equation}
which is the mathematical statement of the temperature inversion symmetry.
It was also shown by Gundersen and Ravndal \cite{G&R88} that the scaled free
energy associated with massless fermions fields at finite temperature
submitted to MIT boundary conditions satisfy the relation given by Eq.({\ref
{TIS}) and therefore exhibts temperature inversion symmetry. Tadaki and
Takagi \cite{TT} have calculated Casimir free energies for a massless scalar
field obeying Dirichlet or Neumann boundary conditions on both plates and
found this symmetry. }For the parallel plate geometry and mixed boundary
conditions it is possible to circunvent the restrictions found by \cite
{R&T89} and discuss temperature inversion symmetry by reconizing that with
respect to the evaluation of the free energy this arrangement is equivalent
to the difference between two Dirichlet (or Neumann) plates \cite{Santosetal}%
. In the case of a massless scalar field at finite temperature and periodic
boundary conditions, it is possible to show that the partition function, and
consequently the free energy, can be written in a closed form such that the
temperature inversion symmetry becomes explicit \cite{CWOT}. Here we shall
show that for the case of the Maxwell field confined by a perfectly
conduting rectangular cavity this symmetry is also present, provided that
we generalize the transformations that relate the high and the low
temperature regimes.

Throughout this letter we employ units such that Boltzmann constant $k_{B}$,
the speed of light $c$ and $\hbar =h/2\pi $ are set equal to the unity.

\section{Evaluation of Helmholtz free energy}

Helmholtz free energy for the confined Maxwell field within a perfectly
conducting rectangular box can be evaluated by means of the generalized zeta
function technique \cite{ZETA}, \cite{Elizaldeetal}. First we introduce the
global zeta function which is defined by 
\begin{equation}
\mu ^{2s}\zeta \left( s\right) =\mu ^{2s}\sum_{p=-\infty }^{\infty
}\,\sum_{\{n\}}\left[ \left( \frac{2\pi p}{\beta }\right) ^{2}+\omega
_{\{n\}}^{2}\right] ^{-s}  \label{GlobalZeta}
\end{equation}
where $\beta =1/T$ , the reciprocal of the temperature $T,$ is the periodic
length along the Euclidean time direction and a mass scale parameter $\mu $
was introduced in order to keep Eq.(\ref{GlobalZeta}) dimensionless; $\omega
_{\{n\}}^{2}$ are the eigenvalues associated with the Euclidean
time-independent modal equation 
\begin{equation}
-\triangle _{\mathbf{x}}\mathbf{\,\,}\varphi _{\{n\}}(\mathbf{x})=\omega
_{\{n\}}^{2}\,\varphi _{\{n\}}(\mathbf{x}),  \label{modaleq}
\end{equation}
where $\mathbf{x}=\left( x,y,z\right) $ and $\triangle _{\mathbf{x}}\mathbf{%
\,}$\ is the Laplacian operator. The Helmholtz free energy function $F\left(
\beta \right) $ for bosonic fields can be obtained from the knowledge of the
global zeta function through the relation 
\begin{equation}
F\,\left( \beta \right) =-\frac{1}{2\beta }\frac{\partial \,}{\partial \,s}%
\left[ \mu ^{2s}\zeta \left( s\right) \right] _{s=0}.  \label{FreeH2}
\end{equation}
The eigenfrequencies associated with the allowed electromagnetic normal
modes within a perfectly conducting rectangular box of linear dimensions $%
a,b,c$ are given by 
\begin{equation}
\omega _{lmn}^{2}=\left( \frac{l\pi }{a}\right) ^{2}+\left( \frac{m\pi }{b}%
\right) ^{2}+\left( \frac{n\pi }{c}\right) ^{2},  \label{eigenfrequencies}
\end{equation}
where $l,m,n$ $\in \{1,2,\ldots \}$. For $l,n,m\neq 0$ there are two
possible polarization states and for $l=0$ or $m=0$ or $n=0$ there is one
polarization state only. Eigenfrequencies for which three or two of the
integers $l,m,n$ are simultaneously zero are not allowed. It follows that
the generalized zeta function for the case in question reads 
\begin{eqnarray}
\mu ^{2s}\zeta \left( s\right)  &=&2\mu ^{2s}\sum_{l,m,n=1}^{\infty
}\sum_{p=-\infty }^{\infty }\left[ \left( \frac{l\pi }{a}\right) ^{2}+\left( 
\frac{m\pi }{b}\right) ^{2}+\left( \frac{n\pi }{c}\right) ^{2}+\left( \frac{%
2p\pi }{\beta }\right) ^{2}\right] ^{-s}  \nonumber \\
&&+\mu ^{2s}\sum_{l,m=1}^{\infty }\sum_{p=-\infty }^{\infty }\left[ \left( 
\frac{l\pi }{a}\right) ^{2}+\left( \frac{m\pi }{b}\right) ^{2}+\left( \frac{%
2p\pi }{\beta }\right) ^{2}\right] ^{-s}  \nonumber \\
&&+\mu ^{2s}\sum_{l,n=1}^{\infty }\sum_{p=-\infty }^{\infty }\left[ \left( 
\frac{l\pi }{a}\right) ^{2}+\left( \frac{n\pi }{c}\right) ^{2}+\left( \frac{%
2p\pi }{\beta }\right) ^{2}\right] ^{-s}  \nonumber \\
&&+\mu ^{2s}\sum_{m,n=1}^{\infty }\sum_{p=-\infty }^{\infty }\left[ \left( 
\frac{m\pi }{b}\right) ^{2}+\left( \frac{n\pi }{c}\right) ^{2}+\left( \frac{%
2p\pi }{\beta }\right) ^{2}\right] ^{-s},  \label{zetaDBox}
\end{eqnarray}
If we now separate the terms corresponding to the zero temperature sector by
setting $p=0$, then it will be easily seen that we can rewrite Eq.(\ref
{zetaDBox}) formally as a sum of Epstein functions (see the Appendix) which
reads 
\begin{eqnarray}
2^{2}\left( \frac{\pi }{\mu }\right) ^{2s}\zeta \,(s) &=&2^{4}E_{4}\left(
s;1/a^{2},1/b^{2},1/c^{2},4/\beta ^{2}\right) +2^{3}E_{3}\left(
s;1/a^{2},1/b^{2},1/c^{2}\right)   \nonumber \\
&&+2^{3}E_{3}\left( s;1/a^{2},1/b^{2},4/\beta ^{2}\right) +2^{2}E_{2}\left(
s;1/a^{2},1/b^{2}\right)   \nonumber \\
&&+2^{3}E_{3}\left( s;1/a^{2},1/c^{2},4/\beta ^{2}\right) +2^{2}E_{2}\left(
s;1/a^{2},1/c^{2}\right)   \nonumber \\
&&+2^{3}E_{3}\left( s;1/b^{2},1/c^{2},4/\beta ^{2}\right) +2^{2}E_{2}\left(
s;1/b^{2},1/c^{2}\right)   \label{zetaDboxtwo}
\end{eqnarray}
In order to regularize Eq.(\ref{zetaDboxtwo}) it is convenient to rewrite it
as a sum of generalized zeta functions and this can be accomplished by
making use of \ Eqs. (\ref{A4}) and (\ref{A5}) in the appendix. Equation (%
\ref{zetaDboxtwo}) then takes the form 
\begin{eqnarray}
2^{2}\left( \frac{\pi }{\mu }\right) ^{2s} &\zeta &\left( s\right)
=A_{4}\left( s;1/a^{2},1/b^{2},1/c^{2},4/\beta ^{2}\right) -A_{2}\left(
s;1/a^{2},4/\beta ^{2}\right)   \nonumber \\
&&-A_{2}\left( s;1/b^{2},4/\beta ^{2}\right) -A_{2}\left( s;1/c^{2},4/\beta
^{2}\right) +4E_{1}\left( s;4/\beta ^{2}\right) .  \label{4DGenZetaTwo}
\end{eqnarray}
Finally, we can regularize Eq.(\ref{4DGenZetaTwo}) by applying the
reflection formula, Eq.(\ref{reflection}) on all terms except the last one
which is already regularized thus obtaining 
\begin{eqnarray}
\mu ^{2s}\zeta \left( s\right)  &=&\frac{\mu ^{2s}abc\beta }{8\pi ^{2}}\frac{%
\Gamma \left( 2-s\right) }{\Gamma \left( s\right) }\,A_{4}\left(
2-s;a^{2},b^{2},c^{2},\beta ^{2}/4\right)   \nonumber \\
&&-\frac{\mu ^{2s}a\beta }{8\pi ^{2}}\frac{\Gamma \left( 1-s\right) }{\Gamma
\left( s\right) }A_{2}\left( 1-s;a^{2},\beta ^{2}/4\right)   \nonumber \\
&&-\frac{\mu ^{2s}b\beta }{8\pi ^{2}}\frac{\Gamma \left( 1-s\right) }{\Gamma
\left( s\right) }A_{2}\left( 1-s;b^{2},\beta ^{2}/4\right)   \nonumber \\
&&-\frac{\mu ^{2s}c\beta }{8\pi ^{2}}\frac{\Gamma \left( 1-s\right) }{\Gamma
\left( s\right) }A_{2}\left( 1-s;c^{2},\beta ^{2}/4\right) +\left( \frac{%
\beta }{2}\right) ^{2s}\zeta _{R}\left( 2z\right) ,  \label{4DGenZetaThree}
\end{eqnarray}
where $\zeta _{R}\left( 2z\right) $ is the Riemann zeta function and we have
also made use of the fact that $E_{1}\left( z;a\right) =a^{-z}\zeta
_{R}\left( 2z\right) $. Notice that all terms on the R.H.S. of Eq.(\ref
{4DGenZetaThree}) are zero when $s\rightarrow 0$ except the last one, that
is, $\zeta \left( 0\right) \neq 0$ since $\zeta _{R}\left( 0\right)
=-(1/2)\ln (2\pi )$. This will give rise to a scale-dependent term absent in
previous calculations related to this problem \cite{Ambjorn&Wolfram}. By
making use of \ Eq.( \ref{FreeH2}) we see that the Helmholtz free energy for
the confined electromagnetic field confined within a perfectly conducting
rectangular box at finite temperature is 
\begin{eqnarray}
F\,\left( a,b,c,\beta \right)  &=&-\frac{abc}{16\pi ^{2}}\sum_{l,m,n,p=-%
\infty }^{\infty }\frac{1}{\left( a^{2}l^{2}+b^{2}m^{2}+c^{2}n^{2}+\frac{%
p^{2}\beta ^{2}}{4}\right) ^{2}}  \nonumber \\
&&+\frac{a}{16\pi }\sum_{l,p=-\infty }^{\infty }\frac{1}{\left( a^{2}l^{2}+%
\frac{p^{2}\beta ^{2}}{4}\right) }+\frac{b}{16\pi }\sum_{m,p=-\infty
}^{\infty }\frac{1}{\left( b^{2}m^{2}+\frac{p^{2}\beta ^{2}}{4}\right) } 
\nonumber \\
&&+\frac{c}{16\pi }\sum_{n,p=-\infty }^{\infty }\frac{1}{\left( c^{2}n^{2}+%
\frac{p^{2}\beta ^{2}}{4}\right) }+\frac{1}{\beta }\ln \left( \frac{\mu
\beta }{\sqrt{2\pi }}\right) .  \label{F}
\end{eqnarray}
The presence of a scale-dependent term is a leftover from the
renormalization procedure carried on via the generalized zeta method and
introduces an ambiguity in the Helmholtz free energy related to the problem
in question. This ambiguity is introduced whenever $\zeta (s)$ is not zero
at $s=0$, see for instance\cite{Elizaldeetal}. It was shown by Myers \cite
{Myers87} that this ambiguty is a legitimate feature of the generalized zeta
function regularization method, whenever the theory is massive or/and
interacting and/or some dimensions are compactified. As consequence the
usual mode sum method and the generalized zeta function method are not
always equivalent. However, the usual mode sum method can be modified so as
to include the complexities of a scale dependence \cite{Blau88}. At zero
temperature and for the simple geometrical cavity that we are considering
here scale-dependent terms are not expected. This is confirmed by the
independent calculations due to Lukosz \cite{Lukosz71}, Ruggiero \textit{et
al} \cite{Ruggieroetal77}, Ambjorn and Wolfram \cite{Ambjorn&Wolfram} and
the present authors. Here it is clear that the scale dependence is
introduced by the generalized zeta function regularization method due to the
imposition of periodic conditions on the Euclidean time direction. For our
purposes the ambiguity in the free energy can be solved by the additional
physical requirement that in unconstrained space, that is, for a very large
rectangular box, the only surviving term must be the Stefan-Boltzmann term.
It is seen then that $\mu $ should be equal to $\sqrt{2\pi}/\beta $.

\section{Temperature inversion symmetry}

The explicit verification is of temperature inversion symmetry in the case
of the system in question is complicated by the fact that in addition to the
inverse temperature parameter $\beta $ we have to deal with three other
length parameters, namely, $a,b$ and $c$, the measures of the sides of the
rectangular cavity. At finite temperature and periodic or antiperiodic
conditions along one spatial direction, or Dirichlet, or Neumann plane
surfaces located perpendicularly to one spatial direction only two
characteristic lengths are involved, a feature that rends this verification
easier. Neverheless, by making recourse to a simple trick we shall show that
temperature inversion symmetry is also present here.

First notice that the denominator of the first term on the R.H.S of Eq.(\ref
{F}) can be rewritten as $a^{2}l^{2}+b^{2}m^{2}+c^{2}n^{2}=\kappa ^{2}\left(
a^{2}q^{2}+b^{2}r^{2}+c^{2}t^{2}\right) $, where $\left\{ q,r,t\right\} $ is
a sequence of three integers with no common factor and $\kappa $ is the
common factor of $\left\{ l,m.n\right\} $. For the sequence $\left\{
q,r,t\right\} $ we define a characteristic length $d_{\left\{ q,r,t\right\}
} $ by 
\begin{equation}
d_{\left\{ q,r,t\right\} }^{2}:=a^{2}q^{2}+b^{2}r^{2}+c^{2}t^{2}.
\end{equation}
Let us also define the dimensionless variable 
\begin{equation}
\xi_{\left\{ q,r,t\right\} }:=\frac{2d_{\left\{ q,r,t\right\} }}{%
\beta }.
\end{equation}
The remaining terms on the R.H.S. of Eq.(\ref{F}) can be treated in a
simpler way. For example, for the second term we define 
\begin{equation}
\xi_{a}:=\frac{2a}{\beta },  \label{chi}
\end{equation}
Then making the replacement: $\sum_{l,m,n=-\infty }\rightarrow
\sum_{\{q,r,t\}}\sum_{\kappa =-\infty }^{+\infty }$, the free energy density
corresponding to Eq.(\ref{F}) can be written as 
\begin{eqnarray}
\frac{F\left( a,b,c,\beta \right) }{abc} &=&-\frac{1}{16\pi ^{2}}%
\sum_{\{q,r,t\}}\sum_{\kappa ,\,p=-\infty }^{\infty }\,\frac{1}{%
d_{\{q,r,t\}}^{4}}\,\frac{\xi_{\left\{ q,r,t\right\} }^{4}}{\left(
\kappa ^{2}\xi_{\left\{ q,r,t\right\} }^{2}+p^{2}\right) ^{2}} 
\nonumber \\
&&+\frac{1}{16\pi a^{2}\,bc}\sum_{l,p=-\infty }^{\infty }\,\frac{\xi%
_{a}^{2}}{\left( l_{a}^{2}\xi_{a}^{2}+p^{2}\right) }  \nonumber \\
&&+\frac{1}{16\pi a\,b^{2}c}\sum_{m,p=-\infty }^{\infty }\,\frac{\xi%
_{b}^{2}}{\left( m_{b}^{2}\xi_{b}^{2}+p^{2}\right) }  \nonumber \\
&&+\frac{1}{16\pi abc^{2}}\sum_{n,p=-\infty }^{\infty }\,\frac{\xi%
_{c}^{2}}{\left( n^{2}\xi_{c}^{2}+p^{2}\right) }
\label{FreeEnergyDensity}
\end{eqnarray}
The zero temperature limit of Eq.(\ref{FreeEnergyDensity}) is obtained by
reconizing that when $\beta \rightarrow \infty $ the dimensionless
parameters $\xi_{\left\{ q,r,t\right\} },\xi_{a},\xi%
_{b},\xi_{c}\rightarrow 0$ and the surviving terms in Eq.(\ref
{FreeEnergyDensity}) correspond to $p=0$, as the reader can easily verify.
Going back to our initial set of indexes $l,m,n$ we obtain 
\begin{equation}
E_{0}==-\frac{abc}{16\pi ^{2}}\sum_{l,m,n=-\infty }^{{+\infty }}\,\frac{1}{%
\left[ a^{2}l^{2}+b^{2}m^{2}+c^{2}n^{2}\right] ^{2}}+\frac{\pi }{48}\left( 
\frac{1}{a}+\frac{1}{b}+\frac{1}{c}\right) ,
\end{equation}
which is in perfect agreement with the result obtained by Lukosz \cite
{Lukosz71} and also by Ruggiero \textit{et al.}\cite{Ruggieroetal77}.

In order to introduce temperature inversion symmetry for this problem let us
now define the dimensionless functions 
\begin{equation}
\mathcal{F}_{\{q,r,t\}}\left( \xi_{\left\{ q,r,t\right\} }\right) :=-%
\frac{1}{16\pi ^{2}}\sum_{\kappa ,\,p=-\infty }^{\infty }\,\,\frac{\mathcal{T%
}_{\left\{ q,r,t\right\} }^{4}}{\left( \kappa ^{2}\xi_{\left\{
q,r,t\right\} }^{2}+p^{2}\right) ^{2}},
\end{equation}
and 
\begin{equation}
\mathcal{F}_{a}\left( \xi_{a}\right) :=\frac{1}{16\pi }%
\sum_{l,p=-\infty }^{\infty }\,\frac{\xi_{a}^{2}}{\left( l_{a}^{2}%
\xi_{a}^{2}+p^{2}\right) },
\end{equation}
with similar definitions for $\mathcal{F}_{b}\left( \xi_{b}\right) $
and $\mathcal{F}_{c}\left( \xi_{c}\right) $. Therefore, in terms of
these functions we can write the free energy density as 
\begin{equation}
\frac{F}{V}=\sum_{\{q,r,t\}}\,\frac{\mathcal{F}_{\{q,r,t\}}\left( \xi%
_{\left\{ q,r,t\right\} }\right) }{d_{\{q,r,t\}}^{4}}+\frac{\mathcal{F}%
_{a}\left( \xi_{a}\right) }{aV}+\frac{\mathcal{F}_{b}\left( \mathcal{%
T}_{b}\right) }{bV}+\frac{\mathcal{F}_{c}\left( \xi_{c}\right) }{cV},
\label{FFreeEnergyDensity}
\end{equation}
where $V=abc$. It is not hard to see that the functions $\mathcal{F}%
_{\{q,r,t\}}\left( \xi_{\left\{ q,r,t\right\} }\right) $ exhibit the
following property 
\begin{equation}
\xi_{\left\{ q,r,t\right\} }^{4}\mathcal{F}_{\{q,r,t\}}\left( \frac{1%
}{\xi_{\left\{ q,r,t\right\} }}\right) =\mathcal{F}%
_{\{q,r,t\}}\left( \xi_{\left\{ q,r,t\right\} }\right) .
\label{delta4}
\end{equation}
In the same way 
\begin{equation}
\xi_{a}^{2}\;f_{a}\left( \frac{1}{\xi_{a}}\right)
=f_{a}\left( \xi_{a}\right) ,  \label{delta2}
\end{equation}
Equations (\ref{delta4}) and (\ref{delta2}) plus two similar equations for
the functions $\mathcal{F}_{b}\left( \xi_{b}\right) $ and $\mathcal{F%
}_{c}\left( \xi_{c}\right) $ describe temperature inversion symmetry
for the case in question, that is, all terms in Eq.(\ref{FFreeEnergyDensity}%
) can be inverted by making use of these formulae.

In the very high temperature limit we expect the leading contribution to be
the Stefan-Boltzmann term, $\pi ^{2}/\left( 45\beta ^{4}\right) $. Now we
argue that our transformations are applied in order to generate from the
unique Stefan-Boltzmann density term, an infinite number of terms which
must be added at zero temperature, and inversely, each term at zero
temperature goes to the unique Stefan-Boltzmann term. In this form, that
is, as far as the energy density is concerned, we can apply the temperature
inversion symmetry transformations to several Casimir systems, including of
course the ones previously analyzed in the literature. The specific form of
the transformations depends on the particular system at hand, but the
procedure is the same.

Now, we can easily check that $\mathcal{F}%
_{\{q,r,t\}}\left( 0\right) =-\pi ^{2}/720$, and that $\mathcal{F}_{a}\left( 
\xi_{a}\right) =\mathcal{F}_{b}\left( \xi_{b}\right) =%
\mathcal{F}_{c}\left( \xi_{c}\right) =\pi /48$. Then, making use of
Eqs.(\ref{delta4}) and (\ref{delta2}) we write 
\begin{equation}
\xi_{\left\{ q,r,t\right\} }^{4}\mathcal{F}_{\{q,r,t\}}\left( \infty
\right) =\mathcal{-}\frac{\pi ^{2}}{720},
\end{equation}
and 
\begin{equation}
\xi_{a}^{2}\;\mathcal{F}_{a}\left( \infty \right) =\xi%
_{b}^{2}\;\mathcal{F}_{b}\left( \infty \right) =\xi_{c}^{2}\;%
\mathcal{F}_{c}\left( \infty \right) =\frac{\pi }{48}.
\end{equation}
Taking these results into Eq.(\ref{FFreeEnergyDensity}) we obtain 
\begin{equation}
F\left( a,b,c,\beta \rightarrow 0\right) \approx -\frac{abc\,\pi ^{2}}{%
45\beta ^{4}}+\frac{\pi }{12\beta ^{2}}\left( a+b+c\right) ,
\end{equation}
which is in agreement with \cite{Ambjorn&Wolfram} with respect to the
leading terms in the high temperature approximation. As expected, the
Stefan-Boltzmann term is the leading term in this limit.

It is convenient to rewrite Eq.(\ref{FreeH2}), or its equivalent Eq.(\ref
{FFreeEnergyDensity}), with the zero temperature terms and the
Stefan-Boltzmann term separated from the non-trivial temperature
corrections, 
\begin{eqnarray}
\frac{F}{V} &=&-\frac{\pi ^{2}}{45\beta ^{4}}+\frac{\pi }{12\beta ^{2}}%
\left( \frac{1}{ab}+\frac{1}{ac}+\frac{1}{bc}\right)  \nonumber \\
&&-\frac{1}{16\pi ^{2}}\sum_{l,m,n=-\infty }^{{+\infty }}\,\frac{1}{\left[
a^{2}l^{2}+b^{2}m^{2}+c^{2}n^{2}\right] ^{2}}+\frac{\pi }{48abc}\left( \frac{%
1}{a}+\frac{1}{b}+\frac{1}{c}\right)  \nonumber \\
&&-\frac{1}{4\pi ^{2}}\sum_{\{q,r,t\}}\sum_{\kappa ,p=1}^{\infty }\frac{1}{%
\left( \kappa ^{2}d_{\{q,r,t\}}^{2}+\beta ^{2}p^{2}\right) ^{2}}+\frac{1}{%
\pi bc}\sum_{l,p=1}^{\infty }\frac{1}{\left( 4a^{2}l^{2}+\beta
^{2}p^{2}\right) }  \nonumber \\
&&+\frac{1}{\pi ac}\sum_{l,p=1}^{\infty }\frac{1}{\left( 4b^{2}l^{2}+\beta
^{2}p^{2}\right) }+\frac{1}{\pi ab}\sum_{l,p=1}^{\infty }\frac{1}{\left(
4c^{2}l^{2}+\beta ^{2}p^{2}\right) }.  \label{FF}
\end{eqnarray}
The last two lines of Eq.(\ref{FF}) represent the non-trivial
temperature-dependent corrections. The first term in the third line is
equivalent to a set of conducting plates with each pair of plates
characterized by a distance $d_{\{q,r,t\}}$, which means that we can take
advantage of results already established in the literature regarding high
and low temperature approximations for a pair of conducting plates.

\section{Conclusions}

In this letter we have employed generalized zeta function techniques in its
global version to derive the finite temperature vacuum energy associated
with an electromagnetic field confined within a perfectly conducting
rectangular box. We have shown that Helmholtz free energy for the problem in
question is scale dependent, a feature that was not present in previous
calculations concerning this problem at zero temperature \cite{Lukosz71}, 
\cite{Ruggieroetal77} and finite temperature \cite{Ambjorn&Wolfram}. The
scale dependence is a remainder that tell us that the theory was
renormalized by the way of the generalized zeta function method. We have
also shown that the concept of temperature inversion symmetry can be
extended so as to include the elctromagnetic field confined by perfectly
conducting rectangular cavity. The main point here is the recognition that
the concept of temperature inversion symmetry can be extended in the sense
that it must be
applied separately to the different terms comprising the non-trivial part
of the free
energy density, with each term behaving differently under this symmetry.
This is the content of Eqs. (\ref{delta4}) and (\ref{delta2}). The most
important feature of temperature inversion symmetry is that it allows us to
establish a relationship between the zero and low-temperature sector of the
Helmholtz free energy and the high-temperature one. This can be very useful
since in general it is easier to obtain low-temperature expansions. In
particular, it is possible to relate the zero temperature Casimir effect to
the Stefan-Boltzmann term in a straightforward way.

\section*{\textbf{Appendix}}

Multidimensional homogeneous Epstein function \cite{Epstein1902} (see also
Elizalde \textit{et al}.\cite{Elizaldeetal} for a modern presentation) are
defined by 
$$
E_{N}\left( z;a_{1},a_{2},...a_{N}\right)
=\sum_{n_{1},n_{2},..n_{N}=1}^{\infty }\left(
a_{1}n_{1}^{2}+a_{2}n_{2}^{2}+\cdots +a_{N}\,n_{N}^{2}\right) ^{-z},
\eqno{(A.1)}  \label{EpsteinE}
$$
for $\Re \,z\,>\,N/2$ and $a_{1},a_{2},...a_{N}\,>\,0$. Generalized zeta
functions. $A_{N}\left( z;a_{1},a_{2},...a_{N}\right) $ are defined by \cite
{Epstein1902},\cite{Elizaldeetal} 
$$
A_{N}\,\left( z;a_{1},a_{2},...a_{N}\right)
=\sum_{n_{1},n_{2},..n_{N}=-\infty }^{+\infty \;\;\prime }\left(
a_{1}n_{1}^{2}+a_{2}\,n_{2}^{2}+\cdots +a_{N}\,n_{N}^{2}\right) ^{-z}, 
\eqno{(A.2)}\label{GenEpsteinFunk}
$$
for $\Re \,z\,>\,N/2$ and $a_{1},a_{2},...a_{N}\,>\,0$. The prime means that
the term corresponding to $n_{1}=n_{2}=...=n_{N}=0$ must be excluded from
the summation.  A useful reflection formula reads envolving these functions
is \cite{Elizaldeetal} 
$$
A_{N}\,\left( z;a_{1},a_{2},...a_{N}\right) =\frac{\pi ^{-\frac{N}{2}+2z}}{%
\sqrt{a_{1}...a_{N}}}\,\frac{\Gamma \left( \frac{N}{2}-z\right) }{\Gamma
\left( z\right) }\,A_{N}\,\left( \frac{N}{2}-z;1/a_{1},1/a_{2},..1/.a_{N}%
\right) . \eqno{(A.3)}\label{reflection}
$$
Relationships between generalized zeta functions and Epstein functions can
be derived from the very definition given by (\ref{GenEpsteinFunk}). For
example, 
$$
A_{2}\left( z;a_{1},a_{2}\right) =2^{2}E_{2}\left( z;a_{1},a_{2}\right)
+2E_{1}\left( z;a_{1}\right) +2E_{1}\left( z;a_{2}\right)
\eqno{(A.4)}\label{A4}
$$
and
 
\begin{eqnarray*}
&&A_{4}\left( z;a_{1},a_{2},a_{3},a_{4}\right) =2^{4}E_{4}\left(
z;a_{1},a_{2},a_{3},a_{4}\right) +2^{3}E_{3}\left( z;a_{1},a_{2},a_{3}\right)
\nonumber \\
&&+2^{3}E_{3}\left( z;a_{1},a_{2},a_{4}\right) +2^{3}E_{3}\left(
z;a_{1},a_{3},a_{4}\right) +2^{2}E_{2}\left( z;a_{1},a_{2}\right)
+2^{2}E_{2}\left( z;a_{1},a_{3}\right)  \nonumber \\
&&+2^{3}E_{3}\left( z;a_{2},a_{3},a_{4}\right) +2^{2}E_{2}\left(
z;a_{2},a_{3}\right) +2^{2}E_{2}\left( z;a_{1},a_{4}\right)
+2^{2}E_{2}\left( z;a_{2},a_{4}\right)  \nonumber \\
&&+2^{2}E_{2}\left( z;a_{3},a_{4}\right) +2E_{1}\left( z;a_{1}\right)
+2E1\left( z;a_{2}\right) +2E1\left( z;a_{3}\right) +2E_{1}\left(
z;a_{4}\right), \hskip 1.4cm (A.5)\label{A5}
\end{eqnarray*}
of which use was made in this letter.

\subsection*{\textbf{Acknowledgments}}

The authors wish to acknowledge useful converstions with Dr. A. A. Actor.

\end{document}